# Microchannel avalanche photodiode with wide linearity range

Z. Sadygov[1,2,*], A. Olshevski[1], N. Anphimov[1], T. Bokova[1], V. Chalyshev[1], I. Chirikov-Zorin[1], A. Dovlatov[2], Z. Krumshtein[1], R. Mekhtieva[2], R. Mukhtarov[2], V. Shukurova[2], M. Troitskaya[1], V. Zhezher[1]

[1] Joint Institute for Nuclear Research, 141980, Dubna, Russia
[2] Institute of Physics, AZ-1143 Baku, Azerbaijan

[*] e-mail: ZSadygov@zecotek.com

Design and physical operation principles of new microchannel avalanche photodiode (MC APD) with gain up to $10^5$ and linearity range improved an order of magnitude compared to known similar devices. A distinctive feature of the new device is a directly biased p-n junction under each pixel which plays role of an individual quenching resistor. This allows increasing pixel density up to 40000 per mm$^2$ and making entire device area sensitive.

In the last decade a significant success was achieved in development of micropixel avalanche photodiodes (MP APD) capable of detecting single photons at room temperature. The MP APD gain can reach values of $10^5$–$10^6$ and photon detection efficiency 20-40 % in a wide wavelength range. It's parameters are comparable with those of vacuum photomultiplying tubes (PMT) [1,2].

Basic design of such devices is described in papers [3, 4]. The device consists of an array of small p-n-junctions (pixels) with typical dimensions 10–100 μm created on the surface of silicon substrate. Pixels are placed with some gap between them in order to avoid charge cross-talk. Each pixel is connected with a common bias line by an individual film resistor with resistance $10^5$–$10^7$ Ω.

Pixel area and resistance of the individual resistor are chosen so that within characteristic time of the pixel capacitance relaxation probability of dark charge carrier generation in the sensitive region was much smaller than one. This provides a possibility for MP APD pixels to operate in overvoltage mode, i.e. at potentials higher than breakdown voltage. A self-quenching avalanche with a constant charge value occurs in the pixel sensitive area then single photoelectron (or dark electron) is created in it. The avalanche is similar to the Geiger discharge. The avalanche in the pixel quenches due to potential drop on it below breakdown voltage. The individual film resistor prevents significant charge of the pixel from power supply during the avalanche process. Signals from fired pixels are added on a common load (bias line) what provides linearity of the device. Photo response of the device is linear until probability that two or more photons hit the same pixel reaches a significant value.

However the above mentioned MP APDs do not have wide enough linearity range because of low pixel density. Design of the device is such that significant part of its surface is occupied by bias line, film resistors, and guard rings at each pixel. For this reason increase of number of pixels over 1000 per $mm^2$ results in significant decrease of sensitive area necessary for photon detection. This in turn leads to decrease of photon detection efficiency (PDE) and as result to decrease of the device amplitude resolution [5].

This paper presents description of design and operation principle of microchannel avalanche photodiode (MC APD) with ultrahigh density of independent multiplication channels that provide for wide photo response linearity range. The distinguishing feature of the MC APD is that it does not have a common bias line and function of an individual quenching resistor is performed by directly biased p-n-junction located under each pixel. Design of the device is shown in Figure 1. The device consists of silicon substrate with n-type of conduction on the surface of which two 4 μm deep epitaxial layers with the same specific resistance of 7 Ω·cm are grown. An array of highly doped regions with $n^+$-type of conduction with a step from 5 to 15 μm, depending on implementation, is formed between the epitaxial layers. This provides for increase in pixel density up to 40000 channels per $mm^2$ and makes entire device surface sensitive. Technology of the MC APD manufacturing is described in [6].

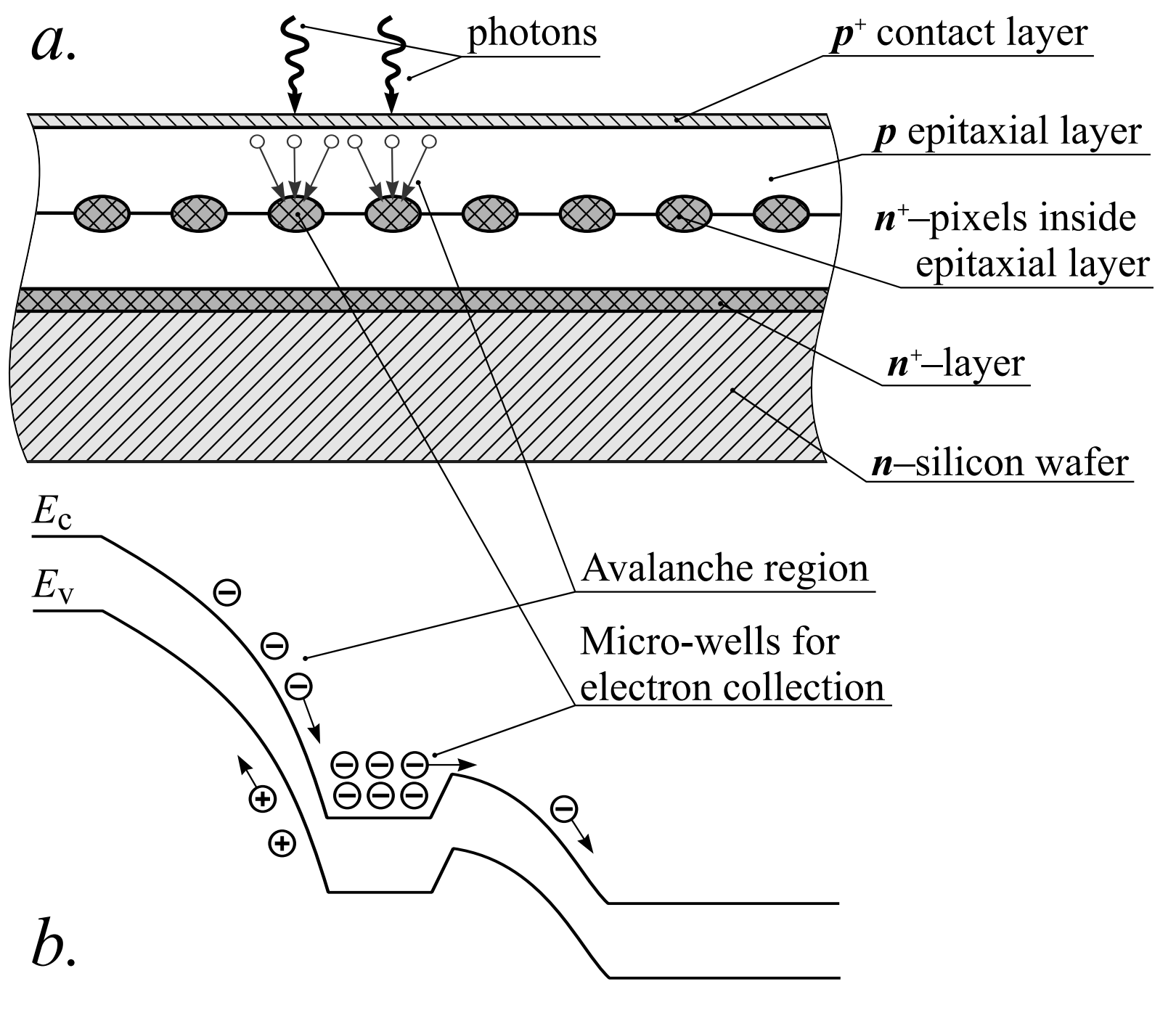

Figure 1. MC APD cross-section (a) and energy zone diagram (b) then bias voltage is applied.

In the operation mode negative relative to the substrate voltage is applied to the MC APD. Depletion of the device starts at the first p-n-junction located between the first epitaxial layer and the substrate. At certain voltage value the depletion reaches the array of $n^+$-regions and partly opens the second p-n-junction located there. After that depletion only the third p-n-junction located between the array of $n^+$-regions and the second epitaxial layer begins. Subsequent increase of the voltage leads to full depletion of the second epitaxial layer. As a result of this an array of potential wells in the $n^+$-regions is initiated in the MC APD depleted region. Above each $n^+$-region there is semispherical electric field that provides for charge collection from the sensitive area of the pixel. Therefore the sensitive area of the device is divided into photosensitive regions with separate multiplication microchannels independent from each other.

Avalanche multiplication of charge carriers takes place in the border region of the second epitaxial layer with $n^+$-regions where high electric field is created. Created in a process of multiplication electrons are accumulated in the potential wells within the $n^+$-regions. This leads to decrease of electric field in the second epitaxial layer below some threshold level. As a result avalanche in given channel quenches. Recovery of the original field in the multiplication microchannel is a result of charge leakage into the substrate through the directly biased p-n-junction between the first epitaxial layer and $n^+$-region.

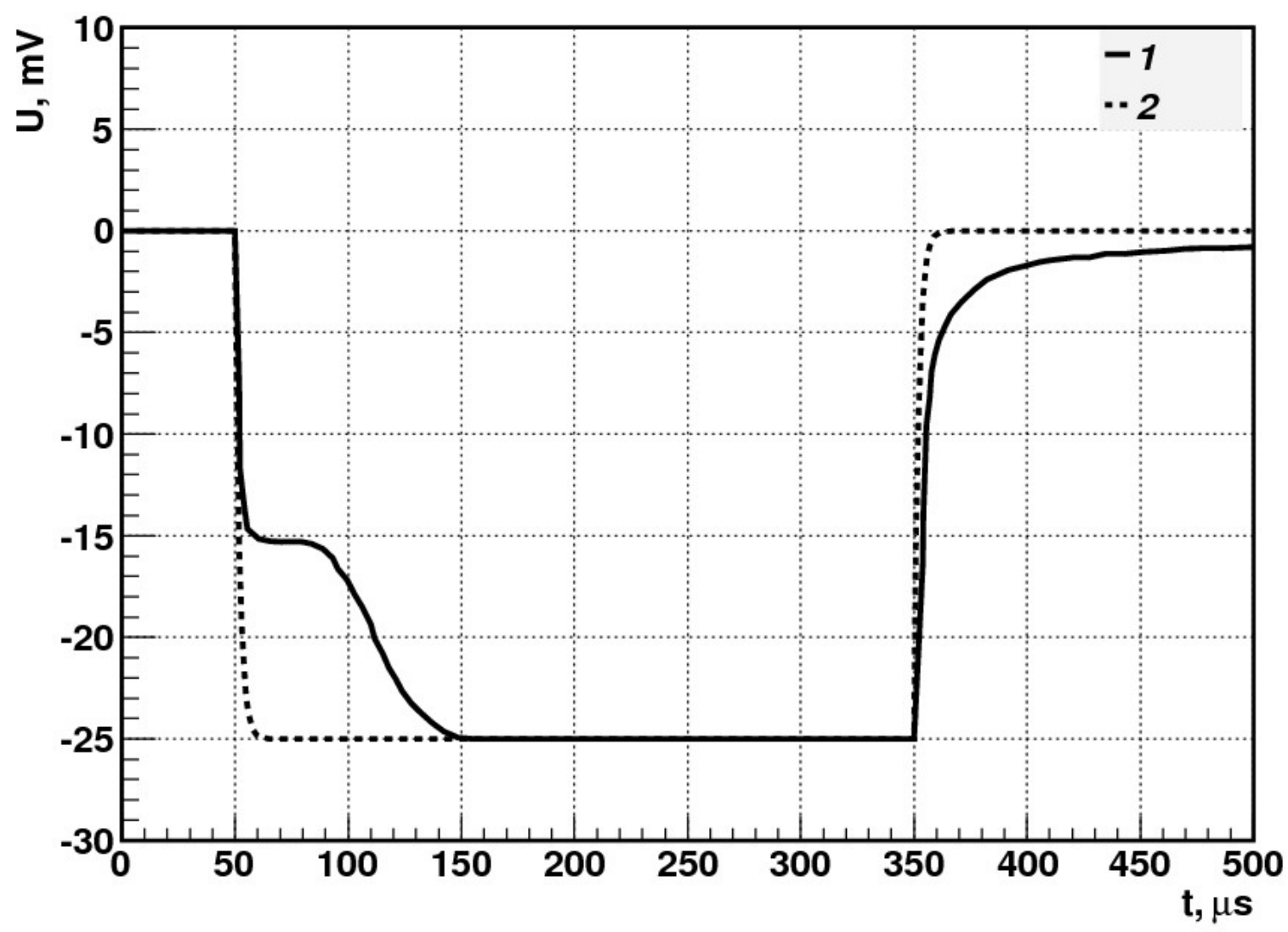

Figure 2. Oscilloscope picture of MC APD (1) and regular APD (2) photoresponse to 300 μs long light pulses. Applied voltage was equal to 20 V and gain was equal to 1.

Photoelectron collection in the $n^+$-regions can be investigated at small bias voltages on the device when there is no amplification of the photoresponse. The MC APD sample under investigation was produced by Zecotek Photonics Inc. and had = $1,35 \cdot 10^5$ multiplication channels on the $3 \times 3$ $mm^2$ sensitive area ($1,5 \cdot 10^4$ channels per $mm^2$). Figure 2 shows an oscilloscope picture of the MC APD photo response to 300 μs long light pulses (curve 1). The signal was read out from 5 kΩ load resistor. For comparison in the same picture a response from regular APD without $n^+$-regions is shown (curve 1). The APD was manufactured parallel to MC APD in the same technological conditions. A light emitting diode (LED) with 450 nm wavelength was used as a light source. Light was completely absorbed in the top part of the second epitaxial layer, i.e. before the depth of the $n^+$-regions location.

In the beginning photoelectrons created photocurrent passing through the second epitaxial layer. Then they were collected in the potential wells causing appearance of a plateau on curve 1 of Figure 2. After some time defined by the light intensity the potential wells were filled and the photoresponse amplitude grew to its maximum which was equal to amplitude of the regular APD. Total charge collected in all potential wells of the device can be determined as an area (integral) of the difference between two curves 1 and 2 in the front part of signals. After graphical integration the total charge $Q_{tot}=1,1 \cdot 10^{-10}$ C was obtained. This means that in each $n^+$-region potential well charge equal to $Q_p=Q_{tot}/N_p=8 \cdot 10^{-16}$ C was collected (about $5 \cdot 10^3$ photoelectrons). This charge provides for electric field screening in the second epitaxial layer, i.e. in the avalanche region. This leads to the avalanche quenching then device operates in avalanche mode.

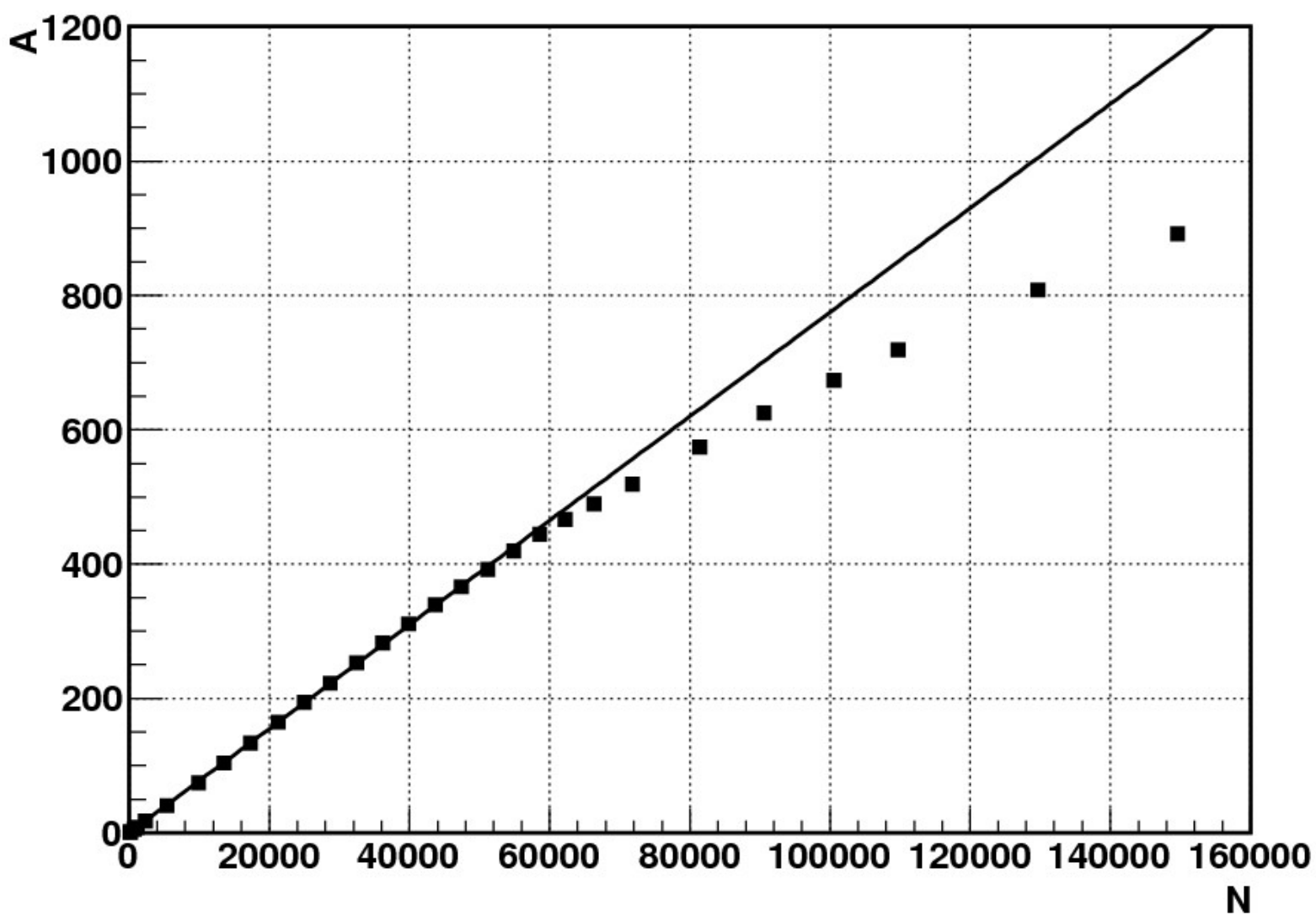

Figure 3. Dependence of MC APD photoresponse amplitude A (in relative units) on a number of incident photons N.

In order to investigate the MC APD photoresponse linearity an operating voltage 90 V was applied to the device. Gain at this voltage was equal to $5 \cdot 10^4$ and photon detection efficiency was equal to 25 %. The device was illuminated with 30 ns long light pulses with 450 nm wavelength and 1 kHz frequency. Hamamatsu S8664-55 APD with known spectral sensitivity was used at gain value equal to one to determine a number of photons in the light pulse. Result of the MC APD photoresponse linearity measurement is shown in Figure 3. One can see that amplitude of the MC APD photoresponse is linear up to $6 \cdot 10^4$ photons in the pulse.

Therefore, new supersensitive avalanche photodiode with high gain and wide photoresponse linearity range was designed and produced. Such a device can be used as light detector in equipment which requires high amplitude resolution, especially in high energy physics and nuclear medicine. The photodiodes were used recently in electromagnetic calorimeter prototypes.